\documentclass[
reprint,
superscriptaddress,
amsmath,amssymb,
aps,
pra,
floatfix,
]{revtex4-2}

\def\final{0}

\usepackage{braket}
\usepackage{mathtools}
\usepackage{graphicx}
\usepackage{color, xcolor}
\usepackage{dcolumn}
\newcolumntype{z}[1]{D{.}{.}{#1}}
\usepackage{physics}
\usepackage{upgreek}
\usepackage{subfigure}
\usepackage{tabularx}
\newcolumntype{Y}{>{\centering\arraybackslash}X}

\usepackage[linesnumbered,ruled ]{algorithm2e}
\SetKwInput{KwOutput}{Output}
\SetKwInOut{Input}{Input}
\SetKwInput{Output}{Output} 
\UseRawInputEncoding 

\usepackage{xargs}
\usepackage[colorinlistoftodos,prependcaption]{todonotes}

\definecolor{Asparagus}{rgb}{0.53, 0.66, 0.42}
\definecolor{cornflowerblue}{rgb}{0.39, 0.58, 0.93}
\definecolor{darkolivegreen}{rgb}{0.33, 0.42, 0.18}
\definecolor{awesome}{rgb}{1.0, 0.13, 0.32}

\newtheorem{definitionenv}{Definition}
\newtheorem{lemmaenv}[definitionenv]{Lemma}
\newtheorem{theoremenv}[definitionenv]{Theorem}
\newtheorem{corollaryenv}[definitionenv]{Corollary}
\newtheorem{propositionenv}[definitionenv]{Proposition}
\newtheorem{conjectureenv}[definitionenv]{Conjecture}
\newtheorem{remarkenv}[definitionenv]{Remark}
\newenvironment{remark}{\begin{remarkenv}\rm}{\end{remarkenv}}
\newcommand{\br}{\begin{remark}}
	\newcommand{\er}{\end{remark}}

\newtheorem{exampleenv}{Example}
\newtheorem{app-lemmaenv}[section]{Lemma}

\newenvironment{definition}{\begin{definitionenv}\rm}{\end{definitionenv}}
\newenvironment{lemma}{\begin{lemmaenv}\rm}{\end{lemmaenv}}
\newenvironment{theorem}{\begin{theoremenv}\rm}{\end{theoremenv}}
\newenvironment{corollary}{\begin{corollaryenv}\rm}{\end{corollaryenv}}
\newenvironment{example}{\begin{exampleenv}\rm}{\end{exampleenv}}
\newenvironment{proposition}{\begin{propositionenv}\rm}{\end{propositionenv}}
\newenvironment{conjecture}{\begin{conjectureenv}\rm}{\end{conjectureenv}}
\newenvironment{app-lemma}{\begin{app-lemmaenv}\rm}{\end{app-lemmaenv}}

\newcommand{\bd}{\begin{definition}}
	\newcommand{\ed}{\end{definition}}
\newcommand{\bl}{\begin{lemma}}
	\newcommand{\el}{\end{lemma}}
\newcommand{\elp}{\hspace*{\fill} $\Box$
\end{lemma}}
\newcommand{\bt}{\begin{theorem}}
\newcommand{\et}{\end{theorem}}
\newcommand{\etp}{\hspace*{\fill} $\Box$
\end{theorem}}
\newcommand{\bc}{\begin{corollary}}
\newcommand{\ec}{\end{corollary}}
\newcommand{\ecp}{\hspace*{\fill} $\Box$
\end{corollary}}
\newcommand{\bcj}{\begin{conjecture}}
\newcommand{\ecj}{\end{conjecture}}

\newcommand{\be}{\begin{example}}
\newcommand{\ee}{\end{example}}
\newcommand{\eep}{\hspace*{\fill} $\Box$
\end{example}}
\newcommand{\bp}{\begin{proposition}}
\newcommand{\ep}{\end{proposition}}
\newcommand{\epp}{
\end{proposition}}

\ifnum\final=0
\newcommand{\mynote}[2]{{\color{#1} \marginpar{\tiny #2}}}
\newcommand{\mybignote}[2]{{\color{#1} $\langle \langle$ #2$\rangle \rangle$}}
\newcommandx{\rednote}[2][1=]{\todo[linecolor=red,backgroundcolor=red!25,bordercolor=red,#1]{#2}}
\newcommandx{\bluenote}[2][1=]{\todo[linecolor=blue,backgroundcolor=blue!25,bordercolor=blue,#1]{#2}}
\newcommandx{\yellownote}[2][1=]{\todo[linecolor=yellow,backgroundcolor=yellow!25,bordercolor=yellow,#1]{#2}}
\newcommandx{\greennote}[2][1=]{\todo[inline,linecolor=olive,backgroundcolor=green!25,bordercolor=olive,#1]{#2}}

\else
\newcommand{\mynote}[2]{}
\newcommand{\mybignote}[2]{}
\newcommand{\rednote}[2][1=]{}
\newcommand{\bluenote}[2][1=]{}
\newcommand{\greennote}[2][1=]{}
\newcommand{\yellownote}[2][1=]{}

\fi

\usepackage{adjustbox}
\usetikzlibrary{quantikz2}
\usetikzlibrary{backgrounds,fit,decorations.pathreplacing}  
\usetikzlibrary{automata} 
\usetikzlibrary{circuits.logic.CDH}
\usetikzlibrary{circuits.logic.US}
\usetikzlibrary{mindmap}
\usetikzlibrary{decorations}
\usetikzlibrary{decorations.pathmorphing}
\usetikzlibrary{arrows,decorations.pathreplacing}

\tikzset{meter/.append style={draw, inner sep=10, rectangle, font=\vphantom{A}, minimum width=30, line width=.4, path picture={\draw[black] ([shift={(.1,.3)}]path picture bounding box.south west) to[bend left=50] ([shift={(-.1,.3)}]path picture bounding box.south east);\draw[black,-latex] ([shift={(0,.1)}]path picture bounding box.south) -- ([shift={(.3,-.1)}]path picture bounding box.north);}}}

\begin{document}

\title{
Quantum keystroke logging
} 

\author{En-Jui Chang}
\email{phyenjui@gmail.com}
\affiliation{Independent researcher, Taichung 421786, Taiwan}

\date{\today}

\begin{abstract}
Superdense coding has long been regarded as a secure quantum communication protocol. It is natural to assume that employing logical quantum states with error-correcting capability would not compromise this security. However, in the context of GKP-based quantum communication, we propose a vulnerability that we term quantum keystroke logging. Specifically, consider a large organization that prepares and transmits logical Bell states without ever assembling the full encoded Bell state. Even under this restriction, the organization can still extract the input information without detection, thereby realizing a form of quantum keystroke logging.
\end{abstract}

\maketitle

\section{Introduction}

Extracting information from a constrained system is a central task across many areas of information processing, though the nature of the information and the relevant constraints vary. Even before distinguishing between classical and quantum settings, two broad scenarios can be identified. (1) In error correction~\cite{PhysRevA.111.052602, hwfz-c6vy,2503.05249}, the goal is to recover information in the form of error syndromes associated with unknown physical events. (2) In secure communication~\cite{PhysRevLett.69.2881}, the goal is instead to prevent unauthorized access to private information. The constraints differ accordingly: in the former, one typically has limited ability to suppress or correct errors, while in the latter, one must assume that an adversary possesses ideal quantum capabilities, including the ability to implement perfect quantum gates. Security cannot rely on the weakness of the adversary.

Although the goals of error correction and secure communication are distinct, both ultimately concern the extraction of information from an unknown operation. In error correction, this operation may be viewed as a stochastic process governed by nature, whereas in communication it corresponds to a deliberate action chosen by the sender. One might argue that logical operations in communication cannot be confused with correctable errors. Yet it remains important to consider whether information could be extracted by means other than standard stabilizer measurements.

This issue becomes more pressing in a realistic scenario where large organizations serve as the sole providers of quantum web-hosting services. Maintaining quantum hardware and establishing reliable long-distance quantum channels are prohibitively expensive, and thus ordinary users may only interact with the system through restricted interfaces. Such users would have limited ability to perform ``private" quantum operations on logical states prepared by the provider. By choosing a particular error-correcting code, ostensibly to enhance reliability, the provider could also constrain the class of available private operations. A natural question then arises: can a quantum operation of a known type but unknown parameter be learned by methods beyond stabilizer measurement, while leaving the logical codewords intact? If not, the system may be secure; if so, the possibility of \textit{quantum keystroke logging} emerges.

Estimating an unknown phase applied to the eigenstate of a given operator, without detailed knowledge of its spectrum, is a standard task known as quantum phase estimation~\cite{kitaev1995}. This motivates the question of whether there exist quantum error-correcting codes in which distinct logical operators are represented as phase shifts. The GKP code~\cite{Gottesman2001} provides such an example, since its logical Pauli operators correspond to displacements in orthogonal phase-space directions.

Two technical challenges arise in adapting phase-estimation techniques to this setting. First, logical GKP codewords are not eigenstates of the logical operators and therefore do not directly reveal the applied phase. Second, in realistic communication scenarios, a logical operation is typically applied only once, rather than many times as required by standard phase-estimation procedures. To address these challenges, we employ an alternative method that extracts the associated geometric phase to overcome the first obstacle, and we introduce a simplified phase-estimation procedure that avoids repeated applications to address the second. Concretely, if a malicious provider were to supply modified GKP states and subsequently infer the logical operations applied by users prior to transmission, \textit{quantum keystroke logging} would become possible.

The techniques we develop rest on three key observations. (1) The geometric phase associated with a closed trajectory in phase space is proportional to the enclosed area and can be detected as an effective Pauli-Z rotation on an ancilla. (2) In standard quantum phase estimation, the role of the quantum Fourier transform (QFT)~\cite{coppersmith1994} is to reshape the probability distribution of measurement outcomes and amplify the likelihood of obtaining the desired result. The essential feature is the probability distribution itself, not the preservation of the detailed superposition structure. This permits the introduction of an auxiliary oscillator to assist the Fourier transform on the ancilla of phase estimation, thereby simplifying the standard QFT. (3) When using oscillators rather than qubits, cross-Kerr nonlinearities~\cite{Venkataraman2012, Ding2017, Heo2021} provide a natural mechanism to reduce the circuit depth required for the QFT.

In this work, we present an example, a sufficient condition, for quantum keystroke logging. Our purpose is not to enumerate all possible vulnerabilities (necessary conditions); the omission of a particular scheme should not be interpreted as evidence of its security. Rather, our aim is to emphasize that the challenges of quantum communication extend beyond error correction, and that security must remain a central consideration in the design of reliable protocols. An open direction is to investigate whether other quantum error-correcting codes, such as the qubit codes~\cite{PhysRevA.111.052602} or the extended binomial codes~\cite{hwfz-c6vy}, are susceptible to similar vulnerabilities or instead avoid them.

\section{Preliminaries}

To establish the necessary background, we begin by revisiting the quantum communication protocol of superdense coding, which is commonly regarded as secure. We then review the GKP code and standard quantum phase estimation, in order to clarify the two technical challenges outlined above as well as the three key observations on which our approach relies.

\subsection{Superdense coding}

Suppose an ordinary sender wishes to transmit two classical bits using the superdense coding protocol provided by an ethical quantum web-hosting service to a receiver. We assume the provider prepares the required entangled pair, sends one qubit to the sender for encoding, and the other qubit directly to the receiver. The sender applies the appropriate operation and returns the encoded qubit to the provider, who forwards it to a base station so that the full entangled pair can be reconstructed at the receiver's site for measurement. The detailed steps are as follows:

\begin{enumerate}
    \item[(i)] The provider prepares a Bell state
    \begin{equation}
        \ket{\Phi^{+}}=\frac{1}{\sqrt{2}}(\ket{0}\ket{0}+\ket{1}\ket{1}).
    \end{equation}
    \item[(ii)] The provider sends the first qubit to the sender for encoding and the second qubit to the receiver. Let the subscripts $S$ and $R$ denote the sender and receiver, respectively. The shared state is then
    \begin{equation*}
        \ket{\Phi^{+}}_{S,R}=\frac{1}{\sqrt{2}}(\ket{0}_S\ket{0}_R+\ket{1}_S\ket{1}_R).
    \end{equation*}
    \item[(iii)] To encode two classical bits $(00, 01, 10, 11)$, the sender applies the operators $(I, X, Z, Y)$ on the received qubit, where $I=\ket{0}\bra{0}+\ket{1}\bra{1}$, $X=\ket{0}\bra{1}+\ket{1}\bra{0}$, $Y=\mathrm{i}(\ket{1}\bra{0}-\ket{0}\bra{1})$, $Z=\ket{0}\bra{0}-\ket{1}\bra{1}$, and $\mathrm{i}=\sqrt{-1}$. The resulting encoded states are:
    \begin{itemize}
        \item[(00)] $\frac{1}{\sqrt{2}}(\ket{0}_S\ket{0}_R+\ket{1}_S\ket{1}_R)$,       
        \item[(01)] $\frac{1}{\sqrt{2}}(\ket{1}_S\ket{0}_R+\ket{0}_S\ket{1}_R)$,
        \item[(10)] $\frac{\mathrm{i}}{\sqrt{2}}(\ket{1}_S\ket{0}_R-\ket{0}_S\ket{1}_R)$,
        \item[(11)] $\frac{1}{\sqrt{2}}(\ket{0}_S\ket{0}_R-\ket{1}_S\ket{1}_R)$.
    \end{itemize}
    \item[(iv)] The sender returns the encoded qubit to the provider. At this stage, the provider does not possess the full entangled pair, since the other qubit already resides at the receiver's site.
    \item[(v)] The provider transmits the encoded qubit to the receiver via the designated base station.
    \item[(vi)] The receiver performs a Bell measurement on the two qubits to infer the transmitted information. The Bell measurement consists of a CNOT gate
    \begin{equation}
        \text{CNOT} = \ket{0}\bra{0}\otimes I + \ket{1}\bra{1}\otimes X,
    \end{equation}
    followed by a Hadamard gate
    \begin{equation}
        H = \frac{1}{\sqrt{2}}(\ket{0}\bra{0}+\ket{0}\bra{1}+\ket{1}\bra{0}-\ket{1}\bra{1}),
    \end{equation}
    and finally two $Z$-basis measurements. Just before measurement, the received states become:
    \begin{itemize}
        \item[(00)] $\ket{0}_R\ket{0}_R$,       
        \item[(01)] $\ket{0}_R\ket{1}_R$,
        \item[(10)] $-\mathrm{i}\ket{1}_R\ket{1}_R$,
        \item[(11)] $\ket{1}_R\ket{0}_R$.
    \end{itemize}
    Thus, the $ZZ$ measurement outcomes $(00,01,10,11)$ correspond to the transmitted classical bits $(00,01,11,10)$, respectively. The apparent crossing in the $01$ and $10$ cases is consistent with the standard superdense coding convention.  
\end{enumerate}

The commonly assumed ``security" of superdense coding relies on the maximally entangled nature of the Bell state. If an eavesdropper intercepts only one qubit, any local measurement yields a maximally mixed state and thus reveals no information. Moreover, the sender encodes only after confirming that the receiver has received their qubit, ensuring that the provider never holds the full pair simultaneously. Nevertheless, as we demonstrate in a later section, information can still be extracted through an alternative strategy, even without simultaneous access to the complete encoded pair.

\subsection{GKP state}

GKP states are bosonic states that encode a qubit into an oscillator by exploiting translational symmetries in phase space, which correspond to the stabilizers of the code. The former perspective is typically favored by physicists, while the latter is more natural in the quantum error-correction community. To make the connection explicit for both audiences, we first relate these two viewpoints before formally defining the GKP code.

The stabilizers of the GKP code are generated by two carefully chosen displacement operators. Each stabilizer commutes with all logical operators (e.g., the logical $X$ and $Z$) as well as with the other stabilizer. For physicists, these are naturally interpreted as translational symmetries in phase space. For the qubit quantum error-correction community, however, the notion of “spanning” stabilizers may appear unconventional, since in standard qubit codes the square of a stabilizer is simply the identity. In fact, the GKP stabilizers can be identified as the squares of the logical Pauli operators, namely $S_X = X^2$ and $S_Z = Z^2$.

To define the logical Pauli operators of the GKP code, we begin with the annihilation operator, proceed to the quadrature operators, and then introduce displacement operators. The annihilation operator is given by $\hat{a}=\sum_{n'=1}^{\infty}\sqrt{n'} \ket{n'-1}\bra{n'},$ with the commutation relation $[\hat{a},\hat{a}^{\dagger}]=1$ in the number basis $\{\ket{n}\}_{n=0}^{\infty}$. The quadratures are defined as the position and momentum operators, $\hat{q} =\frac{\hat{a} +\hat{a}^{\dagger} }{\sqrt{2}}$ and $\hat{p} =\frac{\hat{a}-\hat{a}^{\dagger} }{\mathrm{i}\sqrt{2}},$ which satisfy the canonical commutation relation $[\hat{q},\hat{p}] = \mathrm{i}$. The displacement operator is defined for $\alpha \in \mathbb{C}$ as
\begin{equation}
    \hat{D}(\alpha)= e^{\alpha\hat{a}^{\dagger}-\alpha^{*}\hat{a}}=e^{\sqrt{2}(\Im{\alpha}\hat{q}-\mathrm{i}\Re{\alpha}\hat{p})}.
\end{equation}
In terms of displacement operators, the logical Pauli operators of the GKP code are given by
\begin{equation}    
    X = \hat{D}(\sqrt{\frac{\pi}{2}}),\hspace{1cm} Z = \hat{D}(\mathrm{i}\sqrt{\frac{\pi}{2}}),\hspace{1cm} Y = \mathrm{i}XZ.
\end{equation}

To verify that the stabilizers $S_X$ and $S_Z$ commute with the logical Pauli operators, we apply the Baker–Campbell–Hausdorff (BCH) formula. For general operators $A$ and $B$,
\begin{align}
e^A e^B &= e^C,\
C &= A + B + \tfrac{1}{2}[A,B] + \text{higher-order}.
\end{align}
For two displacement operators $\hat{D}(\alpha)$ and $\hat{D}(\beta)$, let
\begin{align}
    A=&\sqrt{2}(\Im{\alpha}\hat{q}-\mathrm{i}\Re{\alpha}\hat{p}),\notag\\
    B=&\sqrt{2}(\Im{\beta}\hat{q}-\mathrm{i}\Re{\beta}\hat{p}).
\end{align}
Using the canonical commutation relation $[\hat{q},\hat{p}] = \mathrm{i}$ and the fact that scalars commute with all operators, the higher-order nested commutators vanish. Thus,
\begin{equation}
    C =\sqrt{2}(\Im{\alpha+\beta}\hat{q}-\mathrm{i}\Re{\alpha+\beta}\hat{p})+\Im{\alpha\beta^*}.
\end{equation}
Defining the relative phase $\theta=\mathrm{i}\Im{\alpha\beta^*}$, we find that the product of two displacement operators can be written as
\begin{equation}
    e^C = e^{-\mathrm{i}\theta} \hat{D}(\alpha+\beta).
\end{equation}
This formula provides a convenient criterion for determining whether two displacement operators commute or anticommute. Specifically,
\begin{equation}
    \hat{D}(\alpha)\hat{D}(\beta)=e^{-\mathrm{i}\theta}\hat{D}(\alpha+\beta), \hspace{0.2cm} \hat{D}(\beta)\hat{D}(\alpha)=e^{\mathrm{i}\theta}\hat{D}(\alpha+\beta).
\end{equation}
Thus, the operators commute when $2\theta \equiv 0 \pmod{2\pi}$ and anticommute when $2\theta \equiv \pi \pmod{2\pi}$. This phase factor is the essential ingredient in verifying commutation between stabilizers and logical operators and anticommutation among distinct logical operators. More generally, nontrivial values of $\theta$ encode geometric phases, a point to which we will return later.

To contrast with conventional stabilizer codes, recall that a stabilizer set $\mathcal{S}={S_j}$ encoding a qubit into $n$ physical qubits satisfies $S_j^2 = I$. The logical state $\ket{\bar{0}}{\text{qubit}}$ can then be written as
\begin{equation*}
\ket{\bar{0}}_{\text{qubit}} = \bigotimes_{s\in\mathcal{S}} (\frac{I+s}{\sqrt{2}}) \ket{0}^{\otimes n},
\end{equation*}
which is invariant under the action of every stabilizer, since $s(I+s) = s+s^2 = I+s$.

By contrast, in the GKP code the squares of the displacement operators are not the identity but correspond to translational symmetries in phase space. Up to normalization, the logical state $\ket{\bar{0}}$ is therefore defined as
\begin{equation}
\ket{\bar{0}} \propto \sum_{j=-\infty}^{\infty} S_X^j \ket{q=0},
\end{equation}
where $\ket{q=0}$ is the eigenstate of the quadrature operator $\hat{q}$ with eigenvalue $0$. The translational symmetry under $S_X$ is evident in this construction, while that under $S_Z = \hat{D}(\mathrm{i}\sqrt{2\pi})=e^{\mathrm{i}2\sqrt{\pi}\hat{q}}$ is equally manifest.

\subsection{Geometric phase}

Within much of the error-correction community, it is often assumed that once a quantum state can be described in terms of stabilizers and logical operators, any phase arising from the BCH formula is ``trivial." Even when the relative phase takes general values, not restricted to $0$, $\frac{\pi}{2}$, or $\pi$, etc., it is commonly regarded as a negligible global phase. However, this assumption is not always valid. In what follows, we first compare two cases that are frequently considered equivalent up to a global phase, and then demonstrate that the phase in fact corresponds to a geometric phase that can be physically extracted.

Suppose an unknown GKP Pauli operator is denoted by $\hat{D}(\alpha)$. A seemingly ``trivial" way to conjugate this operator with two displacement operators in opposite directions is
\begin{equation}\label{eq:geo_ph}
    \hat{D}(-\beta)\hat{D}(\alpha)\hat{D}(\beta)= e^{-2\mathrm{i}\theta}\hat{D}(\alpha).
\end{equation}
The result differs from the original operator only by a tunable phase factor. Physically, the phase $\theta$ corresponds to the area enclosed by the parallelogram spanned by the two displacement vectors. Thus, any strategy capable of measuring $\theta$ also provides a means to infer the unknown parameter $\alpha$.

\subsection{Standard quantum phase estimation}

As noted above, reducing the problem of learning an unknown GKP Pauli operator to that of measuring an unknown geometric phase naturally motivates the task of phase estimation. Before introducing our specific scheme for extracting information about an unknown logical Pauli operator in the GKP code, we first review the standard quantum phase estimation protocol, which does not fully meet the requirements of our scenario. The detailed steps are as follows:

\begin{enumerate}
    \item[(i)] Prepare $n$ ancilla qubits in the state $\ket{+}^{\otimes n}$ together with the eigenstate $\ket{\psi}$ of a unitary operator $\hat{U}$, satisfying $\hat{U}\ket{\psi}=e^{-2\mathrm{i}\theta}\ket{\psi}$. The joint initial state is
    \begin{equation}
    \ket{+}^{\otimes n}\ket{\psi} = 2^{-\tfrac{n}{2}}\sum_{j=0}^{2^n-1}\ket{j}\ket{\psi},
    \end{equation}
    where $j = j_0 j_1 \dots j_{n-1} \in \{0,1\}^n$ is an $n$-bit binary string and $\ket{j}=\bigotimes_{k=0}^{n-1}\ket{j_k}$.
    \item[(ii)] Apply controlled powers of $\hat{U}$, i.e.,
    \[
    \ket{j}\ket{\psi} \mapsto \ket{j}\hat{U}^j\ket{\psi},
    \]
    where $j$ is interpreted as an integer in binary. The resulting state is
    \begin{equation}
        2^{-\tfrac{n}{2}}\sum_{j=0}^{2^n-1} e^{-2\mathrm{i}j\theta}\ket{j}\ket{\psi}.
    \end{equation}
    \item[(iii)] Apply the QFT on the first $n$ qubits. For computational basis states,
    \begin{equation}
        \text{QFT}_{2^n}\ket{j} = 2^{-\tfrac{n}{2}}\sum_{k=0}^{2^n-1} e^{\tfrac{2\pi \mathrm{i}}{2^n}jk}\ket{k}.
    \end{equation}
    The total state becomes
    \begin{equation}
        2^{-n}\sum_{k=0}^{2^n-1}\sum_{j=0}^{2^n-1} e^{\tfrac{2\pi \mathrm{i}}{2^n}j(k-2^n \tfrac{\theta}{\pi})}\ket{k}\ket{\psi}.
    \end{equation}
    \item[(iv)] To analyze measurement statistics, parameterize the phase as
    \[
    \theta = 2^{-n}\pi(\ell+\delta), \quad \delta \in \big[-\tfrac{1}{2},\tfrac{1}{2}\big),
    \]
    where $\ell \in \{0,\dots,2^n-1\}$. Then the transformed state is
    \begin{equation}
        2^{-n}\sum_{k=0}^{2^n-1}\sum_{j=0}^{2^n-1} e^{\tfrac{2\pi \mathrm{i}}{2^n}j(k-\ell-\delta)}\ket{k}\ket{\psi}.
    \end{equation}
    Hence, the probability of obtaining outcome $\ket{k}$ upon measurement of the first $n$ qubits is
    \begin{equation}\label{eq:prob}
        p(k) = 2^{-2n}\Bigg|\,\sum_{j=0}^{2^n-1} e^{\tfrac{2\pi \mathrm{i}}{2^n}j(k-\ell-\delta)}\Bigg|^2.
    \end{equation}
\end{enumerate}
In the special case $\delta = 0$, the measurement outcome is guaranteed to be $k=\ell$ with probability $p(\ell)=1$, allowing exact inference of $\theta$. For $\delta \neq 0$, the correct outcome $k=\ell$ still occurs with probability bounded below by $\tfrac{4}{\pi^2}$. In practice, this case is irrelevant, since the sender has only four possible inputs $(I, X, Y, Z)$, corresponding to just four possible values of $\theta$. By appropriately choosing $\beta$ as $\pm\sqrt{\pi/2}$ or $\pm\mathrm{i}\sqrt{\pi/2}$, we can ensure that $\theta = \pm\pi/2$ while making $\delta=0$.

Having introduced the standard quantum phase estimation protocol in detail, we now return to the two technical challenges noted earlier: (1) identifying the appropriate eigenstate $\ket{\psi}$ in step (i), and (2) avoiding repeated applications of $\hat{U}$ in step (ii). We address these issues in the following section.

\section{Quantum keystroke logging}

\subsection{Extracting geometric phase via an ancilla}

From Eq.~\ref{eq:geo_ph}, we know how to induce a geometric phase that encodes the desired information. However, we cannot assume that a shifted GKP codeword itself serves as the required eigenstate. Instead, we introduce an ancillary qubit, prepared in a suitable eigenstate, to extract the phase. The procedure is as follows:

\begin{enumerate}
\item[(i)] Prepare an ancillary qubit in a state $\ket{\psi}$ (to be specified below), together with a GKP codeword $\ket{\bar{\phi}}$. The joint initial state is $\ket{\psi}\ket{\bar{\phi}}$.
\item[(ii)] Apply a controlled-displacement operator of the form $\ket{0}\bra{0} \otimes I + \ket{1}\bra{1} \otimes \hat{D}(\beta)$.
\item[(iii)] Apply the logical Pauli operator $\hat{D}(\alpha)$ on the GKP mode.
\item[(iv)] Apply the inverse controlled-displacement $\ket{0}\bra{0} \otimes I + \ket{1}\bra{1} \otimes \hat{D}(-\beta)$. The resulting state is
\begin{equation}
\hat{U}\ket{\psi} \otimes \hat{D}(\alpha)\ket{\bar{\phi}},
\end{equation}
where $\hat{U}=\ket{0}\bra{0} + e^{-2\mathrm{i}\theta}\ket{1}\bra{1}$.
\end{enumerate}

By choosing $\ket{\psi}=\ket{1}$ for the ancillary qubit, step (i) simultaneously provides the initialization required for quantum phase estimation while leaving the logical GKP state unchanged.

\subsection{One-shot quantum phase estimation}

Having established the basic building block for extracting a geometric phase, we now address how to avoid repeated applications of $\hat{U}$. In this setting, we treat the $n$-bit binary string $j$ simply as a positive integer, eliminating the need for $n$ conditional controlled displacements. To achieve this, we employ an ancillary oscillator rather than $n$ ancillary qubits. The controlled powers of $\hat{U}$ are then realized as
\[
\ket{j}_a\ket{1}_b\ket{\bar{\phi}}_c \;\mapsto\; \ket{j}_a \bigl(\hat{U}^j\ket{1}_b\bigr)\ket{\bar{\phi}}_c,
\]
where $j$ is interpreted as an integer (not its binary expansion). Effectively, this conditional operation amplifies $\beta$ to $j\beta$. Using an oscillator for $\ket{j}_a$ makes the extraction of $j$ more natural. The merged protocol proceeds as follows:

\begin{enumerate}
    \item[(i)] Prepare the ancillary oscillator in an equal superposition $2^{-n/2}\sum_{j=0}^{2^n-1}\ket{j}_a$, together with a qubit $\ket{1}_b$ and a GKP codeword $\ket{\bar{\phi}}_c$.
    \item[(ii)] Apply controlled powers of $\hat{U}$:
    \begin{enumerate}
        \item[(a)] Perform the controlled-displacement
        \[
            \sum_{j=0}^{2^n-1} \ket{j}\bra{j}_a \otimes 
            \Bigl(\ket{0}\bra{0}_b \otimes I_c + \ket{1}\bra{1}_b \otimes \hat{D}(j\beta)_c\Bigr).
        \]
        \item[(b)] Apply the logical Pauli operator $\hat{D}(\alpha)$ on the GKP mode.
        \item[(c)] Apply the inverse of the controlled-displacement from step (ii,a).
    \end{enumerate}
    \item[(iii)] Apply the QFT to the ancillary oscillator. Since qubits have been replaced by a single oscillator, the implementation of the QFT can be further simplified; we defer discussion of this simplification until after completing the main description of quantum keystroke logging. For computational basis states,
    \[
        \text{QFT}_{2^n}\ket{j} = 2^{-n/2}\sum_{k=0}^{2^n-1} e^{\tfrac{2\pi \mathrm{i}}{2^n}jk}\ket{k}.
    \]
    The state then becomes
    \[
        2^{-n}\sum_{k=0}^{2^n-1}\sum_{j=0}^{2^n-1} 
        e^{\tfrac{2\pi \mathrm{i}}{2^n}j\bigl(k-2^n \tfrac{\theta}{\pi}\bigr)}
        \ket{k}_a\ket{1}_b\ket{\bar{\phi}}_c.
    \]
    \item[(iv)] For measurement statistics, only the oscillator subsystem matters; the factor $\ket{1}_b\ket{\bar{\phi}}_c$ may be ignored. The resulting probability distribution coincides with Eq.~\ref{eq:prob}.
\end{enumerate}

Note that using either a purely real or purely imaginary $\beta$ in this procedure yields only a single classical bit of information. To fully determine the complex parameter $\alpha$, the procedure can be run in parallel with both real and imaginary choices of $\beta$. This completes the construction of the quantum keystroke logging scheme.

\subsubsection*{Semi-classical bosonic QFT}

We now return to the deferred discussion of simplifying the QFT implementation.  
Let $\hat{n}$ denote the number operator, $\hat{n}\ket{j}=j\ket{j}$ with $j\geq 0$ an integer.  
Consider the two-mode Hamiltonian describing a cross-Kerr nonlinearity,  
\[
\hat{H}_{\textrm{XK}} = \hat{n}_{a} \otimes \hat{n}_{b},
\]
from which we define the unitary
\[
\hat{V}_{2^n} = e^{\left(\tfrac{2\pi\mathrm{i}}{2^n}\, \hat{H}_{\textrm{XK}}\right)}.
\]

Let the two bosonic modes be prepared in the number basis, with mode $a$ as the data mode and mode $b$ as the ancilla:
\begin{subequations}
\begin{align}
    \ket{\psi_a} &= \sum_{j=0}^{\infty} \alpha_{j}\,\ket{j}_a, \\
    \ket{\psi_b} &= 2^{-\tfrac{n}{2}} \sum_{k=0}^{2^n-1} \ket{k}_b.
\end{align}
\end{subequations}
Applying the cross-Kerr interaction gives
\begin{align}
    \hat{V}_{2^n}\bigl(\ket{\psi_a}\otimes\ket{\psi_{b}}\bigr) 
    &= 2^{-\tfrac{n}{2}} \sum_{j=0}^{\infty} \sum_{k=0}^{2^n-1} 
    \alpha_{j}\, e^{\tfrac{2\pi\mathrm{i}}{2^n}jk}\, \ket{j}_a \ket{k}_b.
\end{align}

Thus, the action of $\hat{V}_{2^n}$ maps the bosonic state $\ket{\psi_a}$ into an entangled state with the ancillary oscillator $\ket{\psi_b}$, effectively realizing the bosonic analogue of the QFT.

\subsubsection*{Consistency of the simplified bosonic QFT}

We emphasize that the cross-Kerr interaction does not implement the exact bosonic QFT introduced in step (iii) of the merged protocol. To verify that this simplified version nonetheless suffices, we replace $\text{QFT}_{2^n}$ with the cross-Kerr operation and examine the resulting state and measurement statistics.

After step (iii), the state becomes
\begin{align}
    2^{-n}\sum_{k=0}^{2^n-1}\sum_{j=0}^{2^n-1} 
    e^{\tfrac{2\pi \mathrm{i}}{2^n}j\bigl(k-2^n \tfrac{\theta}{\pi}\bigr)}
    \ket{j}_a \ket{k}_{a'} \ket{1}_b \ket{\bar{\phi}}_c,
\end{align}
where the subscript $a'$ labels the ancillary oscillator introduced by the cross-Kerr scheme.  

Let
\[
\theta = 2^{-n}\pi(\ell+\delta), \quad \delta \in \big[-\tfrac{1}{2},\tfrac{1}{2}\big), \quad \ell \in \{0,\dots,2^n-1\}.
\]
The state can then be rewritten as
\begin{align}
    2^{-n}\sum_{k=0}^{2^n-1}\sum_{j=0}^{2^n-1} 
    e^{\tfrac{2\pi \mathrm{i}}{2^n}j\bigl(k-\ell-\delta\bigr)}
    \ket{j}_a \ket{k}_{a'} \ket{1}_b \ket{\bar{\phi}}_c.
\end{align}

In the ideal case $\delta=0$, measurement of the $a'$ mode yields outcome $\ell$ with certainty. Again, as we have discussed earlier, we can avoid $\delta\neq 0$ by appropriate setting $\beta$.

\section{Conclusion}

In this work, we have demonstrated a concrete example of quantum keystroke logging using GKP-encoded qubits, illustrating that the challenges of quantum communication extend beyond error correction. By exploiting the geometric phase induced by displacement operators, we showed that a malicious provider can infer the logical Pauli operations applied by a user without directly accessing the full entangled pair. Our scheme combines three key elements: (i) extraction of geometric phase via an ancillary qubit, (ii) a one-shot phase estimation protocol using a bosonic ancillary oscillator to avoid repeated applications of the unknown operation, and (iii) a semi-classical bosonic implementation of the quantum Fourier transform using cross-Kerr interactions to simplify the procedure.

This example establishes a sufficient condition for information leakage in GKP-based communication protocols. While it does not exhaustively identify all potential vulnerabilities, it highlights the importance of explicitly accounting for security considerations in the design of quantum communication schemes. Furthermore, our approach demonstrates that even phases often treated as ``trivial" global factors can encode extractable information, suggesting broader implications for other bosonic codes and fault-tolerant architectures. Future work may explore whether similar vulnerabilities exist in alternative codes and how protocol design can mitigate such risks while preserving operational functionality.

\end{document}